\pdfoutput=1
\documentclass[a4paper,11pt,twocolumn]{article}

\usepackage[pdftex,top=2.0cm, inner=1.5cm, outer=1.5cm, bottom=2.0cm]{geometry}	%marginparwidth=4cm,
\usepackage{fancyhdr}
\usepackage{authblk}
\usepackage{abstract}
\usepackage{setspace}
\usepackage{times}
\usepackage{lettrine}
\usepackage{amsmath}
\usepackage{bm}
\usepackage[EULERGREEK]{sansmath}
\usepackage{floatrow}
  \DeclareFloatFont{sansall}{\sffamily\sansmath\small}
\usepackage[font=sansall,labelfont=bf,it]{caption}
\usepackage{graphicx}
\usepackage[usetitle=true,usedoi=true,linkdoi=true]{rsc}	% for submission (reader-friendly)
\setcitestyle{square}
\PassOptionsToPackage{colorlinks=true,linkcolor=magenta,citecolor=green,urlcolor=blue}{hyperref}
\usepackage{hyperref}
\hypersetup{%
pdftitle={Modeling heat dissipation at the nanoscale: An embedding approach for chemical reaction dynamics on metal surfaces},
pdfauthor={Joerg Meyer},
pdfsubject={Angewandte Chemie International Edition Communication},
pdfkeywords={energy dissipation, hot adatoms, embedding, ab initio molecular dynamics, O2 at Pd(100)}
}
\newcommand{\Otwo}{$\text{O}_{\text{2}}$}
\newcommand{\mathid}[1]{ {\text{#1}} }
\newcommand{\idQMper}{\mathrm{\overline{QM}}}
\newcommand{\idMe}{\mathrm{Me}}
\newcommand{\mathset}[1]{ {\mathit{#1}} }
\newcommand{\setR}{\mathset{R}}
\newcommand{\mathvec}[1]{ {\mathbf{#1}} }
\newcommand{\vecR}{\mathvec{R}}
\newcommand{\vecF}{\mathvec{F}}

\begin{document}

% relying on abstract package
%	${TEXMF}/doc/latex/abstract/abstract.pdf
% as e.g. indicated here
%	http://www.tex.ac.uk/cgi-bin/texfaq2html?label=onecolabs
%
\author{J\"org Meyer\thanks{
\href{mailto:joerg.meyer@ch.tum.de}{\nolinkurl{joerg.meyer@ch.tum.de}}\\
\textit{present address:} Leiden Institute of Chemistry, Gorlaeus Laboratories, Leiden University, P.O. Box 9502, 2300 RA Leiden, The Netherlands
}\hspace{0.5em}}
\author{Karsten Reuter} 
\affil{Lehrstuhl f\"ur Theoretische Chemie, Technische Universit\"at M\"unchen,\\ Lichtenbergstra{\ss}e 4, D-85748 Garching, Germany,\\ \url{http://www.th4.ch.tum.de}}
\title{Modeling Heat Dissipation at the Nanoscale:\\ An Embedding Approach for\\ Chemical Reaction Dynamics on Metal Surfaces}
\date{\today}
\twocolumn[
  \begin{minipage}{.55\textwidth}
{\sffamily\Large\textbf{QM/MM for Metals}}
\newline
{\sffamily\small\textbf{Hot or not}: 
An embedding technique for metallic systems makes it possible to model energy dissipation into substrate phonons during surface chemical reactions from first principles. Application to O2 dissociation at Pd(100) predicts translationally ``hot'' oxygen adsorbates as a consequence of the released adsorption energy (ca. 2.6 eV). This questions the instant thermalization of reaction enthalpies generally assumed in heterogeneous catalysis modeling.}
  \end{minipage}
  \hfill
  \begin{minipage}{.4\textwidth} \includegraphics{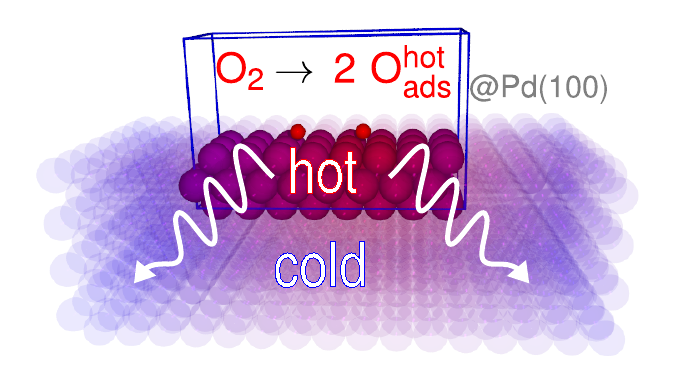}\end{minipage}
  \flushleft
  \maketitle
  \textit{Dedicated to my parents, Brigitte and Siegfried Meyer}
  \vspace{2em}
%   \begin{onecolabstract}\end{onecolabstract}
]
\saythanks
\thispagestyle{fancy}

\paragraph{Abstract:}
\textit{We present an embedding technique for metallic systems that makes it possible to model energy dissipation into substrate phonons during surface chemical reactions from first principles. The separation of chemical and elastic contributions to the interaction potential provides a quantitative description of both electronic \emph{and} phononic band structure. Application to the dissociation of \Otwo\ at Pd(100) predicts translationally ``hot'' oxygen adsorbates as a consequence of the released adsorption energy (ca. 2.6~eV). This finding questions the instant thermalization of reaction enthalpies generally assumed in models of heterogeneous catalysis.}

\renewcommand{\LettrineFontHook}{\fontfamily{pag}\fontseries{bx}\fontshape{it}}
\lettrine[findent=.15em,lines=1]{E}{}xothermic surface chemical reactions may easily release several electron volts of energy. Though sizable in view of potential microscopic dissipation channels, the prevalent picture in chemical kinetics is that this energy is quasi-instantaneously thermalized, ultimately into phononic degrees of freedom. This motivates theoretical treatments on the level of a local temperature and separates the continuous chemical motion into rare-event sequences of thermal reactions. The resulting Markovian state-to-state hopping underlies, for example, all present-day microkinetic formulations in heterogeneous catalysis~\cite{Chorkendorff2003,Reuter2011}. The validation of this picture would require detailed insight into the energy-conversion process at the interface. To date this is limited at best, and if at all centered on reactions at well-defined single-crystal surfaces in ultrahigh vacuum. For a prototypical model reaction like the dissociative adsorption of \Otwo\ molecules scanning-tunneling microscopy experiments have suggested the formation of so-called ``hot adatoms'' on several metal surfaces~\cite{Brune1992,Wintterlin1996,Schintke2001,Yagyu2009}. As a consequence of the released chemical energy, this transient mobility thus intricately couples the elementary reaction steps of dissociation and diffusion. As the experimental quest to generate molecular movies of such reactions is still ongoing, theory has been challenged to elucidate the equilibration dynamics of this process~\cite{Engdahl1994,Ciacchi2004,Gross2009}. Here, the bond breaking and making in highly corrugated surface potentials dictates computationally demanding quantum mechanical (QM) treatments, in particular periodic boundary condition (PBC) supercell approaches to adequately describe the delocalized (surface) metallic band structure \cite{Reuter2011}. Complementing this with a quantitative account of substrate phonons has been cited one of the major conceptual problems in contemporary gas-surface dynamical modeling~\cite{Kroes2008,Tiwari2009,Martin-Gondre2012}. Current state-of-the-art \emph{ab initio} molecular dynamics (AIMD) simulations allow for such substrate mobility, but only within computationally tractable supercells and slab models comprising a few surface lattice constants and layers. In contrast, a simple estimate of the phonon propagation distance in metals according to the speed of sound yields tens of lattice constants per pico second in each direction. Consequently, self-standing AIMD does not provide an accurate reference as unphysical phonon reflections at the PBCs would unavoidably falsify the picosecond-scale adsorbate equilibration dynamics and energy dissipation would be limited to the finite slab thickness. Extending the size of the phononic bath by embedding a QM ``hot reaction zone" into a molecular mechanics (MM) environment described with classical interatomic potentials (CIPs) would be desirable. Unfortunately, QM/MM embedding as routinely employed in biomolecular or materials modeling~\cite{Lin2007,Bo2008,Bernstein2009} is not directly applicable to adsorbate dynamics on \emph{metal (Me)} surfaces, as the introduction of finite QM clusters destroys the proper description of the metallic band structure. In this letter, we present a novel approach coined 'QM/\emph{Me}' that overcomes this limitation by separating the chemical and elastic contributions in the QM interaction potential. By applying this approach to the ``hot adatom'' problem of \Otwo\ dissociation at Pd(100), we can study the equilibration of the dissociation products. By comparison with conventional AIMD, we prove a huge substrate bath to be essential for a correct description of the energy release. Observing a concomitant transient mobility of both O adatom fragments over several lattice constants during the heat dissipation, we thereby provide first-principles support for the experimentally based notion of ``hot adatom''.

\begin{figure*}
\includegraphics[scale=1.0]{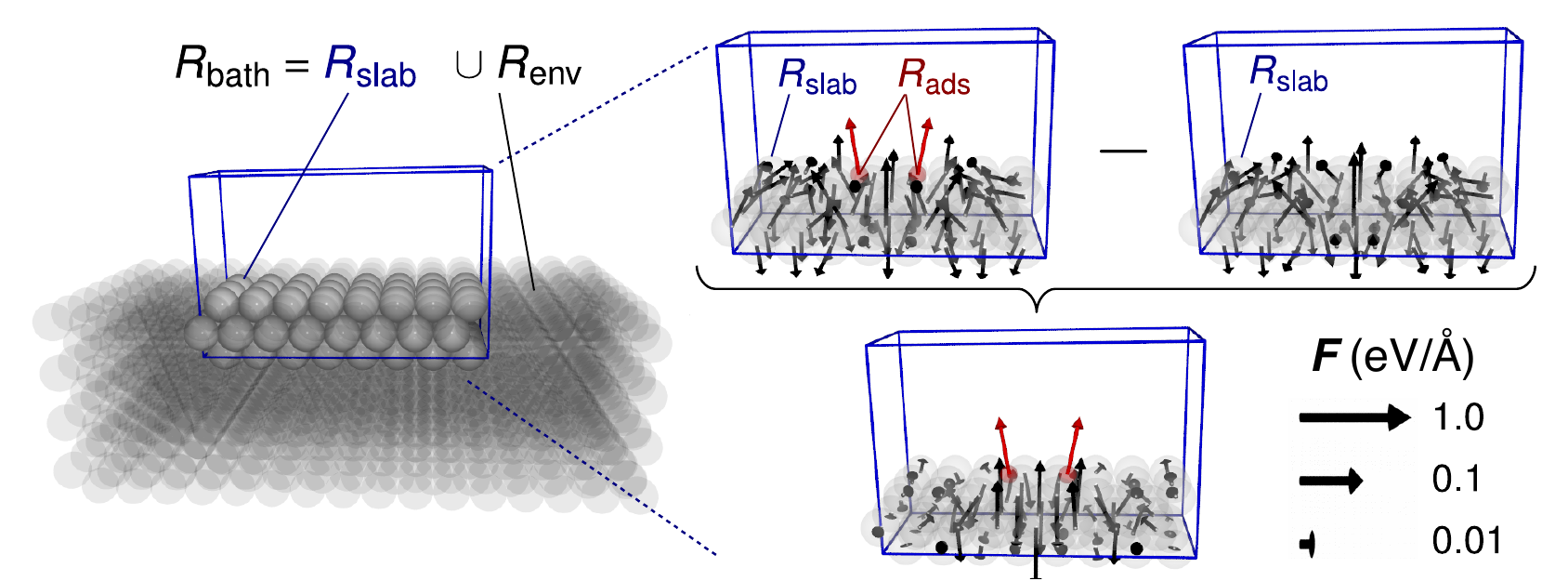}
\caption{\label{fig:QMME_idea}(color online).
Schematic illustration of the QM/Me embedding scheme defined by Equation~(\ref{eq:QMME_V}). A tractable embedding cell for the periodic QM calculations is indicated in blue (3-layer slab formed out of a $3\,\times\,8$ multiple of the primitive surface unit cell). Pd atoms ($\setR_\mathid{slab}$) and the \Otwo\ molecule ($\setR_\mathid{ads}$) contained therein are shown in gray and red, respectively. In the left part, a few additional Pd atoms forming the environment part ($\setR_\mathid{env}$) of the bath ($\setR_\mathid{bath}$) are indicated in transparent black (see text for definitions). In the upper right part, the force fields resulting from the two QM calculations in Equation~(\ref{eq:QMME_V}) acting on Pd and O atoms,   and  , are visualized by black and red arrows, respectively, for a snapshot at $t = 250~\text{fs}$ along the O2@Pd(100) QM/Me-MD trajectory discussed in the text. Resulting forces $\vecF_{I}^{\Delta\idQMper}$ are shown below, illustrating the fast decay of the force differences with increasing distance from the adsorbates that is exploited in QM/Me.}
\end{figure*}

Any embedding scheme tries to improve the description of an interaction locally inside an embedded region. As such it must carefully consider the interaction across the boundary of the latter~\cite{Cai2000,Li2007}, exploiting that the component of the interaction requiring the higher-level description is short-ranged and thus contained within its finite extent. In the case of adsorbate dynamics at metal surfaces, it is the strong chemical adsorbate-substrate and adsorbate-adsorbate interactions that require a higher-level (QM) description so that their complicated many-body nature is correctly accounted for, in particular when bonds are being broken. These components are in principle \cite{Prodan2005} and in practice \cite{Zhang2007} often short-ranged up to exponentially fast decay. As the progressing chemical reaction may lead to lattice deformations, in QM calculations this component is, however, mingled with a second component in form of elastic substrate-substrate interactions. For metallic systems, these elastic interactions are long-ranged as their pairwise contributions typically only slowly decrease as some inverse power of interatomic distance \cite{Prodan2005}. On the other hand, these elastic interactions are already accurately described by many-body CIPs~\cite{Daw1993,Foiles2012}. In QM/Me we thus construct an interatomic potential that disentangles the two types of interactions and implicitly takes care of dealing with the QM/MM boundary:
\begin{eqnarray}
\label{eq:QMME_V}
V^\mathid{QM/Me}(\setR) & = & V^\idMe(\setR_\mathid{bath}) + \\*
  & & \underbrace{ 
    \left[ E^{\idQMper}(\setR_\mathid{slab} \cup \setR_\mathid{ads}) - 
            E^{\idQMper}(\setR_\mathid{slab}) \right] }
    _{ V^{\Delta\idQMper}(\setR_\mathid{slab} \, \cup \, \setR_\mathid{ads}) }
    \; . \nonumber
\end{eqnarray}
where the coordinate sets $\setR$, $\setR_\mathid{bath}$, $\setR_\mathid{slab}$ and $\setR_\mathid{ads}$ are defined as
\begin{subequations}
\label{eq:QMME_coord_sets}
\begin{eqnarray*}
\setR_\mathid{ads} & = & 
  \left\{ \vecR_{X} \,|\, X \in \text{adsorbate atoms in embedding cell} \right\}
  \label{eq:QMME_set_ads} \\
\setR_\mathid{slab} & = &
  \left\{ \vecR_{M} \,|\, M \in \text{metal atoms in embedding cell} \right\}
  \label{eq:QMME_set_slab} \\
\setR_\mathid{env} & = &
  \left\{ \vecR_{M} \,|\, M \in \text{metal atoms in environment} \right\}
  \label{eq:QMME_set_env} \\
\setR_\mathid{bath} & = &
  \left\{ \vecR_{M} \,|\, M \in \text{bath} \right\} = \setR_\mathid{slab} \cup \setR_\mathid{env}
  \label{eq:QMME_set_bath} \\
\setR & = & 
  \left\{ \vecR_{I} \,|\, I \in \text{model} \right\} = \setR_\mathid{bath} \cup \setR_\mathid{ads}
  \label{eq:QMME_set_model}
  \> .
\end{eqnarray*}
\end{subequations}
Both Equations~(\ref{eq:QMME_V}) and the definitions given by Equations~(\ref{eq:QMME_coord_sets}) are illustrated in Figure~\ref{fig:QMME_idea}. $V^{\Delta\idQMper}$ is obtained from two QM calculations within identical PBC supercells defined by what we henceforth term the \emph{embedding cell}: One QM calculation includes the adsorbate atoms while the other does not, and the positions of the substrate atoms are identical in both cases. 

By forming the difference of the two QM calculations, contributions of elastic interactions in the metal substrate are canceled, leaving the specific chemical adsorbate-substrate and adsorbate-adsorbate interactions intended to be described on a QM level. The contribution of elastic substrate-substrate interaction is supplied by $V^\idMe$, which denotes an energy from a large bath-like MM region described at the level of many-body CIPs. Equation~(\ref{eq:QMME_V}) bears similarities to the ONIOM scheme of Morokuma and coworkers~\cite{Svensson1996}, but avoids the construction of finite QM clusters entirely by treating the embedded region twice at the same level of theory. This crucial difference exploiting the short-rangedness of $V^{\Delta\idQMper}$ avoids boundary effects and yields a quantum mechanical many-body augmentation of the CIP $V^\idMe$ that fully captures the chemistry of bond breaking and making.

We illustrate the QM/Me embedding scheme with the application to the dissociative adsorption of \Otwo\ at Pd(100), where ``hot adatom'' motion has also been suspected \cite{Chang1987,Liu2004,Vattuone2010,Liu2013}. While the competition of phononic and electronic dissipation channels during such adsorbate dynamics is currently under heavy debate~\cite{Juaristi2008,Luntz2009,Juaristi2009,Martin-Gondre2012}, our previous work for this specific system has allowed us to obtain a first-principles based estimate of the energy loss into electron-hole pair excitations of less than 5\% of the total chemisorption energy~\cite{Meyer2011}. In the following, we therefore focus on an accurate description of the phononic dissipation channel as a very important, if not dominant part of the dissipation dynamics. Nevertheless, we note that this by no means excludes future use of the interaction potential given by Equation~(\ref{eq:QMME_V}) in an electronically non-adiabatic context. Within an approach based on density-functional theory (DFT) and neural network interpolation~\cite{Goikoetxea2012}, we first analyze the initial gas-surface dynamics up to distances from the surface where the frozen-surface approximation is still valid. As we will describe in detail elsewhere~\cite{Meyer2012diss}, this shows that the surface potential steers essentially all \Otwo\ molecules in particular with thermally distributed impingement velocities into one extremely dominant entrance channel, in which they approach the surface side-on and with their center of mass above a fourfold hollow site. For the present context this thus defines suitable initial conditions for a QM/Me-based molecular dynamics (MD) trajectory evaluating the ensuing dissociation dynamics and energy dissipation into the mobile substrate.

\begin{figure}
\includegraphics[scale=1.0]{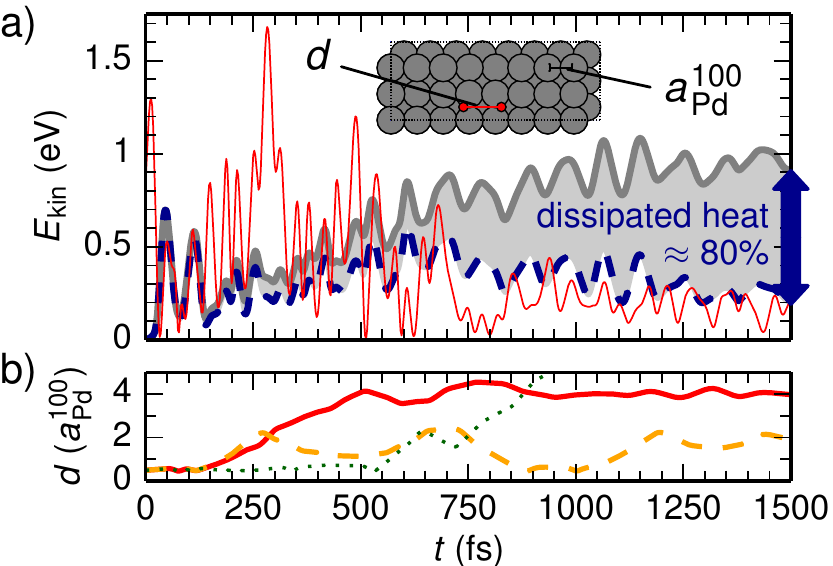}
\caption{\label{fig:QMME_results}(color online). 
a) Energy dissipation during the \Otwo{}@Pd(100) QM/Me-MD trajectory based on the embedding cell shown in Figure~\ref{fig:QMME_idea}, which is centered in a bath of 125,000 atoms described by $V^\idMe$ ($3\,\times\,8$ multiple of the primitive surface unit cell, 3 layers). 
The kinetic energy of the adsorbates $E_{\text{kin}}(\dot{\setR}_{\text{ads}})$ (thin red line), of all Pd atoms $E_{\text{kin}}(\dot{\setR}_{\text{bath}})$ (thick gray line) and of the those in the QM/Me embedding cell $E_{\text{kin}}(\dot{\setR}_{\text{slab}})$ (dashed blue line) are plotted as a function of time. $E_{\text{kin}}(\dot{\setR}_{\text{bath}}) - E_{\text{kin}}(\dot{\setR}_{\text{slab}})$ (gray) is a measure for the dissipated heat. We emphasize that the total energy is well conserved within our QM/Me embedding scheme and provide a detailed discussion in the supporting material. 
b) Separation distance $d$ of the O atoms in units of the surface lattice constant $a_\mathrm{Pd}^{100}$ (see top view inset in (a)) as obtained with QM/Me (solid red line) as well as AIMD with a mobile (dashed orange line) and a frozen surface (dotted green line).}
\end{figure}

The effective separation of short-ranged chemical and long-ranged elastic interactions achieved in QM/Me, and the concomitant fast decay of the difference forces  $\vecF_{I}^{\Delta\idQMper}$ towards the cell boundaries is demonstrated in the right part of Figure~\ref{fig:QMME_idea}. A quantitative analysis is provided in the supporting material, together with technical details of our implementation. While this validates the QM/Me idea, Figure~\ref{fig:QMME_results}~ (a) underscores the importance of the properly described heat dissipation into the bath: About 1.5~ps after the initial \Otwo\ bond dissociation phonons have propagated about 80\% of initially released 2.6\,eV chemisorption energy (at DFT-PBE level) outside the embedding cell. In other words, conventional AIMD simulations based on such a supercell alone would fail to capture the vast majority of the dissipated heat already on that time scale. This also has important consequences for the actual adsorbate dynamics. Figure~\ref{fig:QMME_results}~(b) shows that the motion of the two oxygen adatoms immediately after the dissociation of their parent \Otwo\ molecule is characterized by a largely increasing distance, that is, indeed a ``hot adatom'' motion over several Pd surface unit cells. Starting from the identical initial conditions and using the QM/Me embedding cell as PBC supercell for conventional AIMD trajectories obtained within the same MD setup yield significantly different results also plotted in Figure~\ref{fig:QMME_results}(b) for both a mobile and a frozen surface. The ``hot adatom'' motion is thus intricately influenced by the description of the energy uptake and dissipation into the substrate. Clearly, the thermalization process is not instantaneous on the time scale of the actual adsorbate dynamics. According to the barrier of 300\,meV (DFT-PBE) or even more \cite{Chang1987,Liu2004,Liu2013} for the hollow-bridge-hollow diffusion path followed by the O adatoms, transition state theory would estimate individual thermal diffusive hops at room temperature to take place over time scales of microseconds or even longer. Instead, we observe a transient mobility over several lattice constants within 1.5\,ps. Standard Markovian state-to-state dynamics decoupling dissociation and diffusion thus introduces an error of more than six orders of magnitude. Should reactions with other adsorbates in the vicinity of the \Otwo\ impingement point also be stimulated this way, further paradigm shifts would be required to accommodate such ``hot chemistry'', for example, in our current understanding of heterogeneous catalysis. QM/Me will enable to address these aspects in systematic future studies, where then also acquisition of trajectory statistics using dynamically adapted embedding cells will play a key role. We envision QM/Me to soon become a useful addition to prevalent Langevin approaches to (gas-)surface dynamics \cite{Adelman1977,Tully1981,Kroes2008,Martin-Gondre2012} since it is coming at essentially the same computational cost as AIMD as far as relevant for such sampling: it will allow to validate heat bath assumptions therein and provide a first-principles route to determining their effective materials parameters. A detailed atomistic understanding of the influence of surface temperature and underlying phonon dynamics is thus finally in reach.

In summary, we have presented the novel QM/Me embedding approach that overcomes the inapplicability of conventional QM/MM embedding to metallic systems. Our application to the highly exothermic \Otwo\ dissociation on Pd(100) has made it possible to investigate the dissipation of the adsorption energy from first-principles for the first time, and predicts the formation of ``hot adatoms'' for this system. This paves the way for a deeper understanding of the complex interaction dynamics of adsorbates with phonons beyond the harmonic approximation and under non-equilibrium conditions -- thus allowing to gauge assumptions about energy sinks made in model Hamiltonians used to describe such kind of non-equilibrium dynamics. Within a multi-scale modeling philosophy, this offers an interesting perspective of a more detailed atomistic understanding of energy conversion at interfaces in general. In addition to adsorbate dynamics, QM/Me is generally suitable for application to any problem that involves long-range elastic effects and breaks translational symmetry in a metallic system, for example, also bulk defects. We envision its central idea of how to achieve an effective localization of adsorbate--substrate interactions also to be stimulating and beneficial for the development of future many-body interatomic potentials.

{\sffamily\sansmath\small
}

\paragraph{Acknowledgements:}
Funding by the Deutsche Forschungsgemeinschaft is gratefully acknowledged, as is generous CPU time at the Leibniz Rechenzentrum der Bayerischen Akademie der Wissenschaften.

\bibliographystyle{angew}
\bibliography{qmme}

\end{document}